# Mixing Kohonen Algorithm, Markov Switching Model and Detection of Multiple Change-Points: An Application to Monetary History

Marie-Thérèse Boyer-Xambeu[1], Ghislain Deleplace[2], Patrice Gaubert[3], Lucien Gillard[4] and Madalina Olteanu[5]

1. Université de Paris VII – LED mtb-xambeu@wanadoo.fr

2. Université de Paris VIII – LED ghislain.deleplace@univ-paris8.fr

3. Université de Paris XII – ERUDITE Patrice.Gaubert@univ-paris12.fr

4. CNRS – LED lucien.gillard@wanadoo.fr

5. Université de Paris I – CES SAMOS Madalina.Olteanu@univ-paris1.fr

**Abstract.**
The present paper aims at locating the breakings of the integration process of an international system observed during about 50 years in the 19th century. A historical study could link them to special events, which operated as exogenous shocks on this process. The indicator of integration used is the spread between the highest and the lowest among the London, Hamburg and Paris gold-silver prices. Three algorithms are combined to study this integration: a periodization obtained with the SOM algorithm is confronted to the estimation of a two-regime Markov switching model, in order to give an interpretation of the changes of regime; in the same time change-points are identified over the whole period providing a more precise interpretation of the various types of regulation.

## 1. Introduction

Previous research aimed at combining Kohonen algorithms and Markov switching models to suggest a periodization of the international bimetallism in the 19th century (Boyer-Xambeu, Deleplace, Gaubert, Gillard and Olteanu, 2006). This research was based on an economic study of the international monetary system ruling at this time in Europe, which combined three monetary zones: a gold-standard one, centred in London, a bimetallic one, centred in Paris, and a silver-standard one, centred in Hamburg (Boyer-Xambeu, Deleplace and Gillard, 2006). The three major financial centres of that system (London, Paris, and Hamburg, hence the label LPH used hereafter) were linked through arbitrage operations between markets for gold and silver and markets for foreign exchange located in those centres. Since two metals, gold and silver, acted as monetary standards in that system, it worked as an international bimetallism. Its growing integration during half a century (from 1821 to 1873) was reflected in the convergence of the observed levels of the relative price of gold to silver in London,

Paris, and Hamburg. However, this integration process was subject to various changes, which can be understood as exogenous shocks disturbing that process.
One such shock is vastly documented in the literature: the discovery of new gold mines in the United States and Australia, which led to a sudden decline in 1850 of the gold-silver price over all the markets in the world. This decline was not of the same magnitude everywhere, and therefore the spread between the London, Paris, and Hamburg gold-silver prices increased, stopping for a time the integration process of the system. This is what we will call a breaking in that process. The present paper aims at locating the major breakings occurring during the period of international bimetallism; a historical study could link them to special events, which operated as exogenous shocks on that system. The indicator of integration used is the spread between the highest and the lowest among the London, Paris, and Hamburg gold-silver prices.
Three algorithms are combined to study this integration: a periodization obtained with the SOM algorithm is confronted to the estimation of a two-regime Markov switching model, in order to give an interpretation of the changes of regime; at the same time change-points are identified over the whole period providing a more precise interpretation of these varying types of regulation.
Section 2 summarizes the results obtained with the SOM algorithm to differentiate the sub-periods obtained using the whole available data.
Section 3 presents the kind of model used and the results of its estimation using the new indicator, the spread computed at each period of quotation between the three relative prices of gold in silver. The sub-periods are confronted to the two regimes obtained and some evidence of a relation between the regime and the volatility of the spread is presented.
Section 4 presents the technique used to identify change-points in the temporal process and some strong results of breaks in mean and in variance of the spread are obtained. They are interpreted in terms of monetary history as, for some of them, they are quite new in the literature of this domain.
Some further directions of research are indicated in conclusion.

## 2. The sub-periods obtained with a SOM algorithm[1].

### 2.1 The data

The relative prices of gold in silver are computed from the price of each metal observed, twice a week, in each of the three financial places, Paris, London and Hamburg (respectively, *poa*, *lgs*, and *hoa*), from the beginning of 1821 until the end of 1860. The same type of data is available for the exchange rates (Pound in Francs, Pound in Marks, Mark in Francs: respectively, *lpv*, *hlv*, and *phv*).
An observation is a set of twelve values, two quotations (Tuesday and Friday) for each of the six variables.

[1] Details may be found in Boyer-Xambeu,...,Olteanu,2006.

A computed variable has been added to emphasize the relation between the relative price of metals in Hamburg and the average level in Paris and London of this value (*hpl*).
Most of the time the quotations show rather small differences within a given week, but periods with important troubles, Paris in the late 1840s for instance, may be well separated from the more classical ones.
After the Kohonen classification using a grid of 25 nodes, a hierarchical ascending classification is used to produce a small number of macro classes, in this case 6 macro classes, corresponding to the main sub-periods. This latter classification is constructed with the code vectors obtained from the first process.

## 2.2 Characteristics of the macro-classes

Large sequences of contiguous weeks are grouped in the macro-classes, however a few years are fragmented in short periods situated in different classes

- Class 1 is constituted of 3 groups of years 1829-30, 1834-38, 1848-49 and a lot of fragments of other years
- Class 2 is more simple to describe with 3 intervals 1832-33, 1842-43 and 1846-47 and some sparse weeks from the 1830s.

They represent a central position contrasting to the well identified other classes:

- Class 3: 2 sets constituted of years 1824-25 and 1827-28, with almost no missing weeks in these intervals, indicating that this sub-period is very homogeneous
- Class 4: the end of year 1853 and the whole period 1854-60; again only a small number of weeks are missing for this continuous sub-period of more than seven years
- Class 5: 1821-24 and 1826-beginning 1827 plus small parts of 1830 and 1832
- Class 6: two sets 1839-41 and 1851-53

The means of the variables used to obtain the classification can be represented to illustrate the great differences appearing between the sub-periods. Changing hierarchies between the relative prices are the characteristic identifying the four last macro-classes.

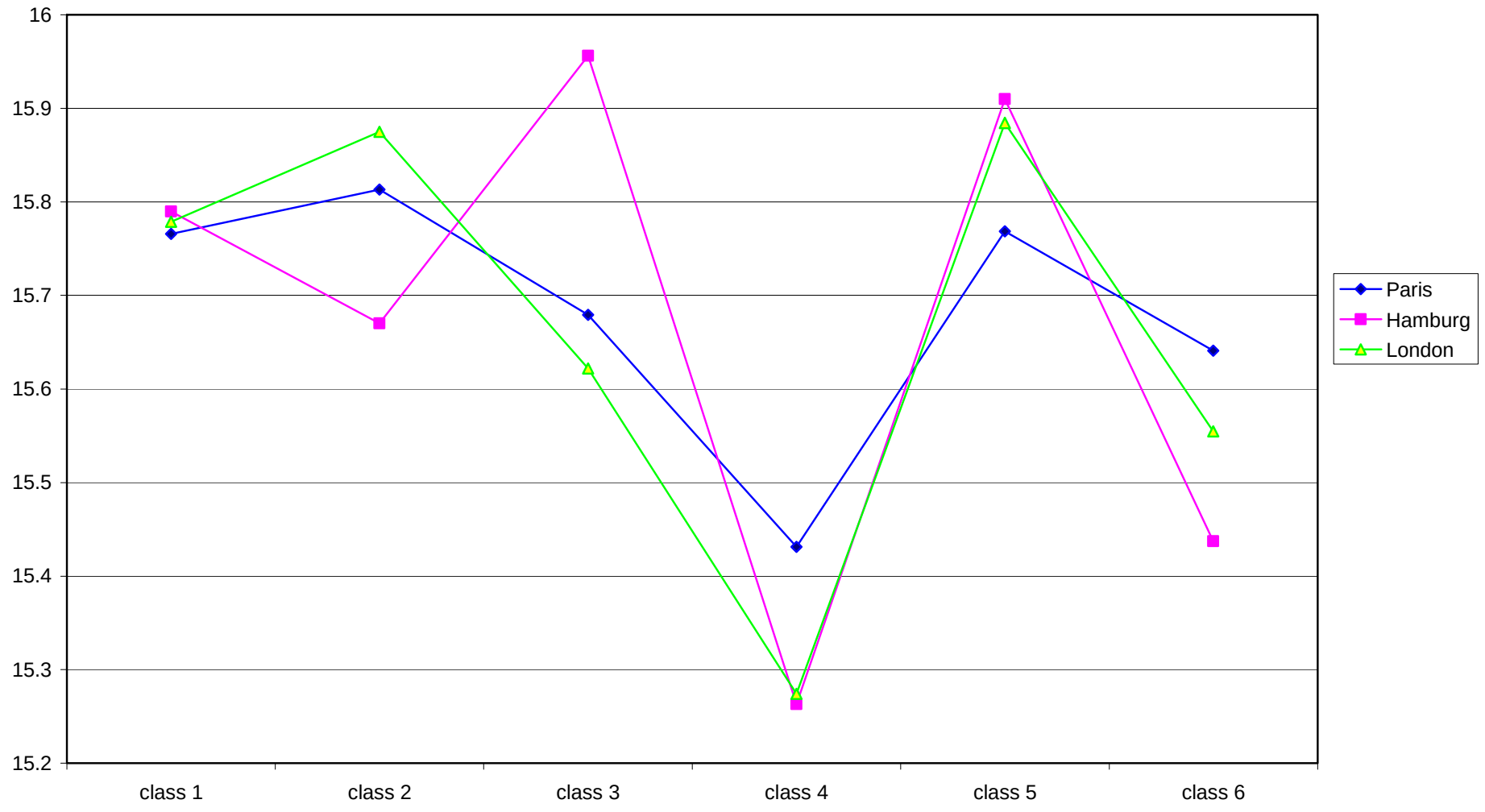

**Fig. 1.** Gold-silver price and the 6 macro-classes

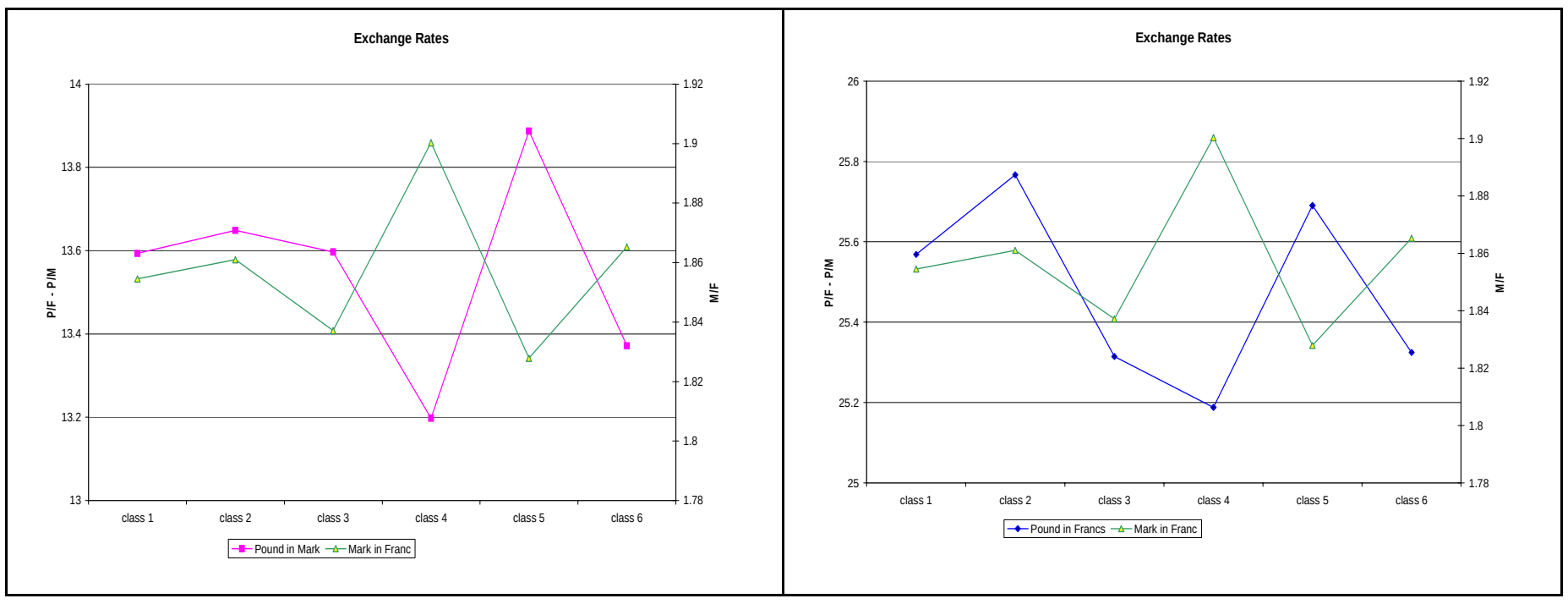

**Fig. 2.** Exchange rates and the 6 macro-classes

Rearranging the various classes according to calendar time allows to distinguish between three sub-periods: a) the 1820s (classes 5 and 3, covering 1821 to 1828); b) the 1830s and 1840s (classes 1 and 2, covering 1829 to 1849); c) the 1850s (classes 6 and 4, covering 1851 to 1860).

Only the years 1839-41 resist to that rearrangement, since they belong to class 6, while they should appear in classes 1 and 2 relative to the 1830s and 1840s; some explanation will be suggested in the last section.

Fig. 1. exhibits two contrasted situations, where the gold-silver price is respectively low (class 4) and high (class 5) in all the three financial centres. Fig. 2. confirms that opposition, since the two classes are also sharply contrasted by the levels of the exchange rates. Years 1821-23 and 1826 (class 5) are marked by a low mark/franc ex-

change rate and high gold-silver prices, the Hamburg one being higher than the Paris one; years 1854-60 (class 4) are marked by a high mark/franc exchange rate and low gold-silver prices, the Hamburg one being below the Paris one.
These remarks, which also apply respectively to the rest of the 1820s (class 3) and to the rest of the 1850s (class 6) are consistent with historical analysis: while the Hamburg mark was always anchored to silver, the French franc was during the 1820s and 1850s anchored to gold (in contrast with the 1830s and 1840s when it was anchored to silver); it is then normal that the mark depreciated against the franc when silver depreciated against gold, and more in Hamburg than in Paris (as in class 5 and 3), and that the mark appreciated against the franc when silver appreciated against gold, and more in Hamburg than in Paris (as in class 4 and 6).

## 3. A model for the spread between the highest and the lowest gold-silver price.

### 3.1 An autoregressive Markov switching model

The key assumption is that the time series to be modeled follow a different pattern or a different model according to some unobserved, finite valued process. Usually, the unobserved process is a Markov chain whose states are called "regimes", while the observed series follows a linear autoregressive model whose coefficients depend on the current regime.
Let us put this in a mathematical language. Suppose that $(y_t)_{t\in Z}$ is the observed time series and that the unobserved process $(x_t)_{t\in Z}$ is a two-states Markov chain with transition matrix

$$A = \begin{pmatrix} p & 1-q \\ 1-p & q \end{pmatrix}, \text{ where } p, q \in \left]0,1\right[ \qquad (1)$$

Then, assuming that $y_t$ depends on the first $l$ lags of time, we have the following equation of the model:

$$y_t = a_0^{x_t} + a_1^{x_t} y_{t-1} + \ldots + a_l^{x_t} y_{t-l} + \sigma^{x_t} \varepsilon_t \qquad (2)$$

where $a_i^{x_t} \in \{a_i^1, a_i^2\} \in R^2$ for every $i \in \{0,1,\ldots,l\}$, $\sigma^{x_t} \in \{\sigma^1, \sigma^2\} \in (R_+^*)^2$ and $\varepsilon_t$ is a standard Gaussian noise.

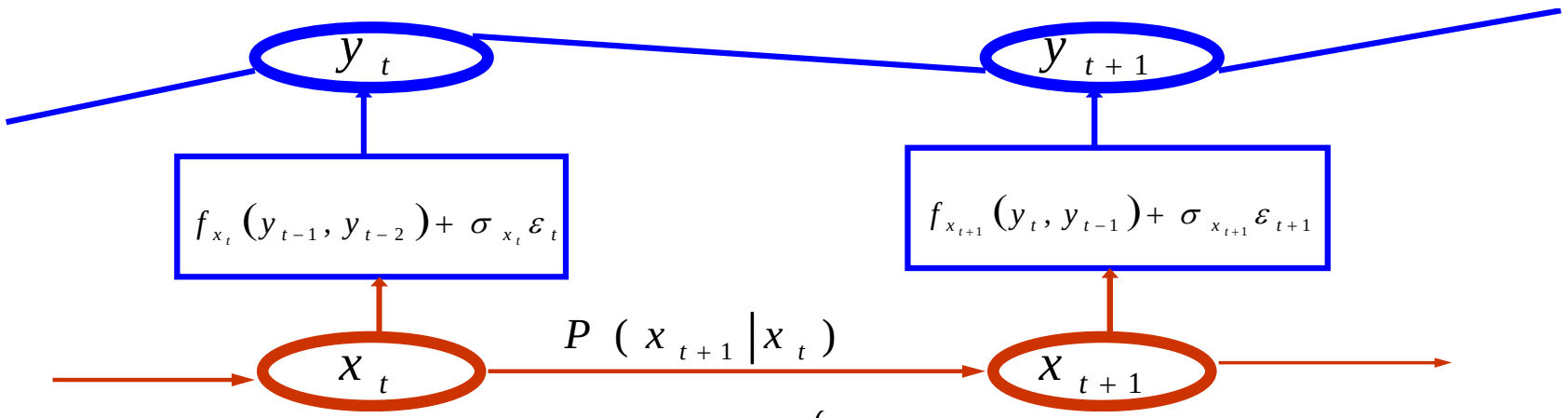

The parameters of the model are then $\{a_0^1, a_1^1, ..., a_l^1, a_0^2, a_1^2, ..., a_l^2, \sigma^1, \sigma^2, p, q\}$ and they are usually estimated by maximizing the log-likelihood function via an EM (Expectation – Maximization) algorithm.
Our characteristic of interest will be the “a posteriori” computed conditional probabilities of belonging to the first or to the second regime. Indeed, as our goal is to derive a periodization of the international bimetallism, the “a posteriori” computed states of the unobserved Markov chain will provide a natural one.
Although the results obtained with a switching Markov model are usually satisfying in terms of prediction and the periodizations are interesting and easily interpretable, a difficulty remains: how does one choose the number of regimes? In the absence of a complete theoretical answer, the criteria for selecting the “right” number of regimes are quite subjective from a statistical point of view.

### 3.2 The results

In this paper we use a two-regime model to represent the spread computed with the gold-silver prices observed at each period on the three places. The transition matrix indicates good properties of stability:

$$\begin{pmatrix} 0.844298 & 0.253357 \\ 0.155702 & 0.746643 \end{pmatrix}$$

and no three regime model was found with an acceptable stability.
The first regime is a multilayer perceptron with one hidden layer, the second one is a simple linear model with one lag. Using the probabilities computed for each regime at each period, it may be interesting to study the six sub-periods obtained and to observe the switch between the regimes along these periods of time.
Most of the time the regime 1 explains the spread (about 70% of the whole period) but important differences are to be noted between the sub-periods:

**Table 1.** Regime 1 and volatility of spread

| Sub-periods | Number of obs. | % regime 1 | Standard deviation of spread |
|---|---|---|---|
| 1 | 483 | 0.733 | 0.053 |
| 2 | 335 | 0.627 | 0.061 |
| 3 | 191 | 0.445 | 0.075 |
| 4 | 376 | 0.816 | 0.044 |
| 5 | 303 | 0.625 | 0.050 |
| 6 | 390 | 0.723 | 0.049 |

Classes 3 and 4 clearly contrast with, respectively, the highest and the lowest volatility of spread as they are ruled by, respectively, regime 2 and regime 1 models.
As will be explained later, further investigations have to be made with a more complex model and using a more adapted indicator of the arbitrages ruling the markets.

## 4. Identification of change-points: a global vision of the bimetallist system of payments.

### 4.1 Elements about the technique[2]

A different approach to model changes of regime in a time-series is to detect change-points or breaks. Here, the main assumption is that the whole series is observed and change-points are computed “a posteriori”. Thus, this approach has not a predictive goal, but it is rather aimed at explaining the series by a piecewise stationary process which seems to be well adapted to our problem.
Mathematically, the model can be written as follows: let us consider the observed *m*-dimensional series $y_t = (y_{1,t},\ldots,y_{m,t})^T$, $t = 1,\ldots,T$ and suppose that it is abruptly changed. The changes, whose number and configuration are unknown, occur in the marginal distribution and may be in mean, in variance or in both mean and variance. We assume that there exists an integer $K^*$ and a sequence of change-points $\tau^* = \{\tau_1^*,\ldots,\tau_{K^*}^*\}$ with $\tau_0^* = 0 < \tau_1^* < \ldots < \tau_{K^*-1}^* < \tau_{K^*}^* = T$ such that $(\mu_k, \Sigma_k) \neq (\mu_{k+1}, \Sigma_{k+1})$ where $\mu_k = E(Y_t)$ and $\Sigma_k = Cov(Y_t) = E(Y_t - E(Y_t))(Y_t - E(Y_t))^T$, $\tau_{k-1}^* + 1 \leq t \leq \tau_k^*$.

The numbers of changes as well as their configuration are computed by minimizing a penalized contrast function. Details on the algorithms for computing the change-points configuration $\tau^*$ can be found in Lavielle and Teyssière (2005).

### 4.2 Some results and interpretation

Applying this technique to the spread gave 7 change-points in mean and 4 in mean and variance.
Fig. 3 summarizes the spread, the four change-points (the first 4 green lines in chronological order) obtained in mean and variance, and the 2 last change-points in mean which correspond to a major break in the level of the gold-silver price, observed simultaneously on the three places and correspond to the great change in production of gold in United States.

---

[2] The authors are very grateful to Gilles Teyssière for a significant help on this part.

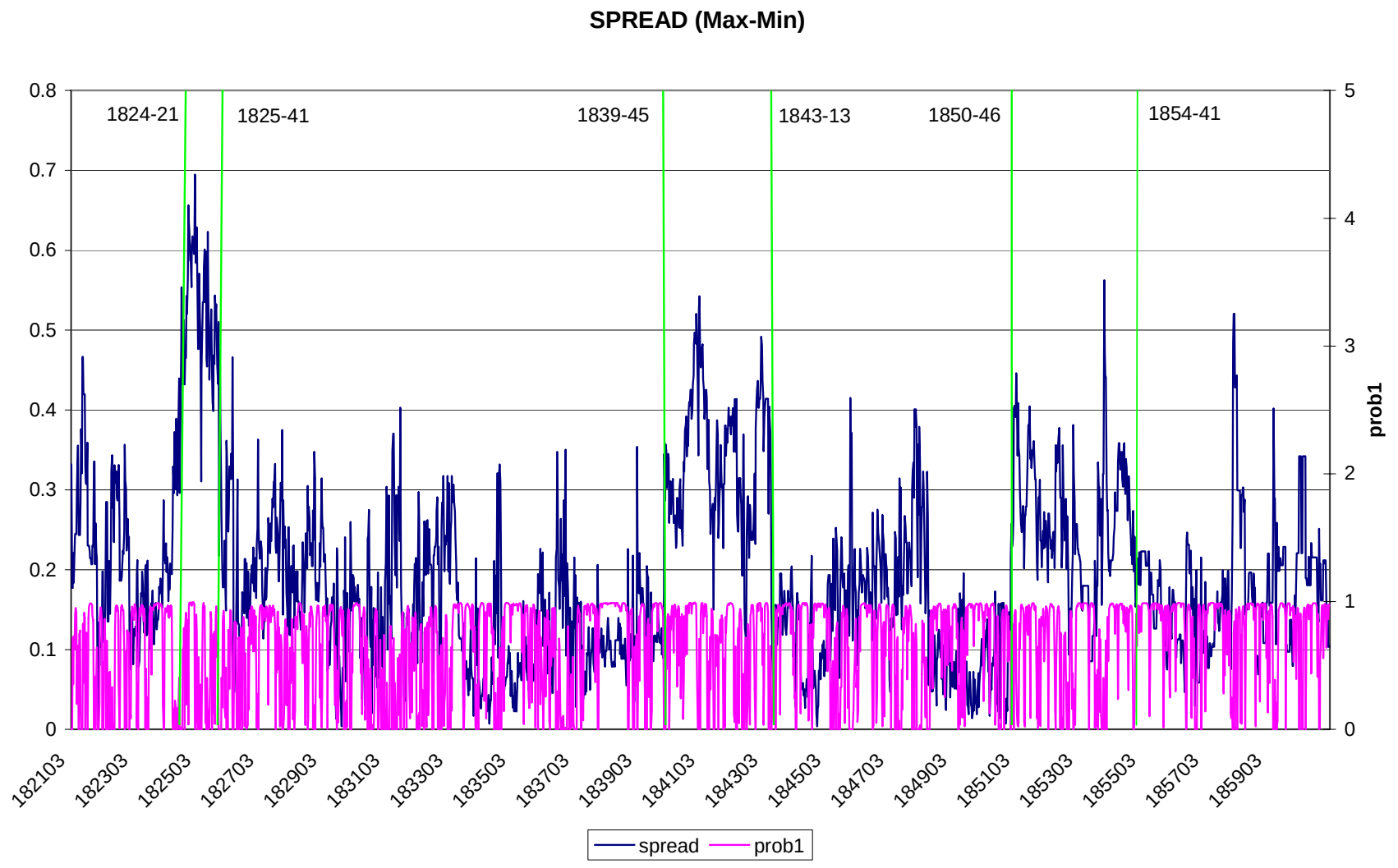

**Fig. 3.** Spread, change-points and probability of regime 1

A closer look at the spread between the highest and the lowest among the London, Hamburg and Paris gold-silver prices draws attention upon three episodes, each of them beginning with a break which sharply increases the spread and ends with another breaking which sharply narrows it (green vertical lines on Fig. 3). These episodes have in common to be linked to shocks affecting the integration process of the LPH system, although the shocks may have been asymmetrical (only one or two of the financial centres being initially hit) or symmetrical (the three of them being simultaneously hit).

The first episode runs from the 21st week of 1824 till the 41st week of 1825. The sharp initial increase in the spread may be explained by two opposite movements in London and Hamburg: on one side, heavy speculation in South-American bonds and Indian cotton fuelled in London the demand for foreign payments in silver, which resulted in a great increase in the price of silver and a corresponding decline in the gold-silver price; on the other side, the price of gold rose in Hamburg while the price of silver remained constant, sparkling the huge spread between the highest (Hamburg) and the lowest (London) gold-silver prices. More than one year later, the opposite movements took place: the price of gold plunged in Hamburg, while the price of silver remained at its height in London, under the influence of continuing speculation (which would end up in the famous banking crisis of December 1825); consequently the spread abruptly narrowed, this event being reflected by the breaking of the 41st week of 1825.

The second episode runs from the 45th week of 1839 till the 13th week of 1843. It started with the attempt of Prussia to unify the numerous German-speaking independent states in a common monetary zone, on a silver standard. This attempt resulted in an increasing demand for silver in central and northern Europe, which put an upward pressure on its price. This created a shock on Hamburg, which was then the main financial centre connecting that zone with western Europe. Since the Bank of Hamburg maintained the price of silver fixed, that pressure led to a drop in the Hamburg price of gold, and consequently in its gold-silver price, at a time when it was more or less stabilized in Paris. The spread between the highest (Paris) and the lowest (Hamburg) gold-silver price suddenly was enlarged, and during more than three years remained at a level significantly higher than during the 14 preceding years. This episode ended with the breaking of the 13$^{th}$ week of 1843, when, this shock having been absorbed, the gold-silver price in Hamburg went back in line with the price in the two other financial centres.

The third episode runs from the 46$^{th}$ week of 1850 till the 41$^{st}$ week of 1854. The shock was then symmetrical: London, Paris and Hamburg were hit by the pouring of gold following the discovery of Californian mines, and the sudden downward pressure on the world price of that metal. But the impact of that pressure differed in each centre, according to the ruling monetary regime: in London, where the sterling was on a fixed gold standard, the price of silver went up; conversely in Hamburg, where the mark was on a fixed silver standard, the price of gold went down; and in Paris, where the franc was on a bimetallic standard, the Bank of France tried to smooth the variations of both prices. This differential impact increased the spread, which had remained near zero during two years, the highest (Paris) gold-silver price falling far less than the lowest (Hamburg). It took four years to absorb this enormous shock, as reflected by the breaking of the 41$^{st}$ week of 1854.

## Conclusion

In the three cases, the integration process of the LPH system, shown by the downward trend of the spread over half a century, was jeopardized by a shock: a speculative one in 1824, an institutional one in 1839, a technological one in 1850. But the effects of these shocks were absorbed after some time, thanks to active arbitrage operations between the three financial centres of the system. Generally, that arbitrage did not imply the barter of gold for silver but the coupling of a foreign exchange operation (on bills of exchange) with the transport of one metal only.

As a consequence, it would be appropriate in a further study to locate the breakings of another indicator of integration: the spread between a representative "national" gold-silver price and an arbitrated international gold-silver price taking into account the foreign exchange rates. At the same time it would be interesting to go further with the Markov switching model, trying more complete specifications.